\documentclass[a4paper,twoside]{article}
\usepackage[authoryear]{natbib}
\bibpunct{(}{)}{;}{a}{}{,}
\usepackage{graphicx}
\usepackage{amssymb}
\usepackage{verbatim}
\usepackage{amsmath} 
\date{} 
\title{\large\bf\flushleft Expanding Space: the Root of all Evil?\thanks{Research undertaken as part of the Commonwealth Cosmology
Initiative (CCI: www.thecci.org), an international collaboration
supported by the Australian Research Council}}
\author{\parbox{\textwidth}{\flushleft
\vspace{-0.5cm}
{\it Matthew J. Francis$^{1,4}$, Luke A. Barnes$^{1,2}$, J. Berian James$^{1,3}$
\& Geraint F. Lewis$^1$}\\
\vspace{0.4cm}
{\small \,$^1$Institute of Astronomy, School of Physics, University of Sydney, NSW 2006, Australia}\\ 
{\small \,$^2$Institute of Astronomy, Madingley Rd, Cambridge, UK}\\
{\small \,$^3$Institute for Astronomy, Blackford Hill, Edinburgh EH9 3HJ, U.K.}\\
{\small \,$^4$Email: {\tt mfrancis@physics.usyd.edu.au}}}}
\begin{document}
\twocolumn[
\begin{changemargin}{.8cm}{.5cm}
\begin{minipage}{.9\textwidth}
\vspace{-1cm}
\maketitle
%
%
\small{{\bf Abstract:}  While it remains  the staple of  virtually all
  cosmological teaching, the concept  of expanding space in explaining
  the increasing  separation of galaxies has recently  come under fire
  as a  dangerous idea whose  application leads to the  development of
  confusion and the establishment of misconceptions. In this paper, we
  develop a  notion of expanding space  that is completely  valid as a
  framework for the  description of the evolution of  the universe and
  whose application allows an intuitive understanding of the influence
  of universal  expansion.  We also demonstrate  how arguments against
  the concept in general have failed thus far, as they imbue expanding
  space with physical properties  not consistent with the expectations
  of general relativity. }



\medskip
\medskip
\end{minipage}
\end{changemargin}
]
\small
\section{Introduction} \label{intro} 

When  the mathematical  picture of  cosmology is  first  introduced to
students in senior undergraduate or junior postgraduate courses, a key
concept to be  grasped is the relation between  the observation of the
redshift  of galaxies  and  the general  relativistic  picture of  the
expansion of the Universe.  When presenting these new ideas, lecturers
and textbooks often resort to analogies of stretching rubber sheets or
cooking raisin bread  to allow students to visualise  how galaxies are
moved apart,  and waves of light  are stretched by  the ``expansion of
space''.  These kinds of analogies are apparently thought to be useful
in giving students a mental picture of cosmology, before they have the
ability to directly comprehend  the implications of the formal general
relativistic description.  However,  the academic argument surrounding
the  expansion of  space  is  not as  clear  as standard  explanations
suggest; an  interested student and  reader of New Scientist  may have
seen \citet{ns} state
\begin{quote}
  ...how is it possible for  space, which is utterly empty, to expand?
  How   can  nothing   expand?   The  answer   is:   space  does   not
  expand. Cosmologists sometimes talk  about expanding space, but they
  should know better.
\end{quote}
while being told by \citet{2000cosm.book.....H} that
\begin{quote}
  expansion redshifts  are produced by the expansion  of space between
  bodies that are stationary in space
\end{quote}
What is a  lay-person or proto-cosmologist to make  of this apparently
contradictory situation?

Whether or not  attempting to describe the observations  of the cosmos
in  terms of  expanding  space is  a  useful goal,  regardless of  the
devices used to do so,  is far from uncontroversial. Recent attacks on
the physical concept of expanding  space have centred on the motion of
test particles in the expanding universe; \citet{2004Obs...124..174W},
\citet{Peacockweb}  and others  claim  that expanding  space fails  to
adequately  explain the  motion of  test particles  and hence  that it
should  be abandoned.   But what,  exactly, is  at  fault?  Crucially,
these claims rely on  falsifying predictions made from using expanding
space as  a tool to guide  intuition, to bypass  the full mathematical
calculation.  However, the very  meaning of the phrase expanding space
is not rigorously defined, despite  its widespread use in teaching and
textbooks.  Hence,  it is prudent to  be wary of  predictions based on
such a poorly defined intuitive frameworks.

In  recent work,  \citet{barnes2006} solved  the test  particle motion
problem for universes with  arbitrary asymptotic equation of state $w$
and found agreement between  the general relativistic solution and the
expected behaviour  of particles in expanding space.   We suggest that
the  apparent conflict  between  this work  and  others, for  instance
\citet{Chodconformal}, lies predominantly in differing meanings of the
very concept of expanding space.   This is unsurprising, given that it
is  a phrase  and concept  often stated  but seldom  defined  with any
rigour.

In this  paper, we examine the  picture of expanding  space within the
framework  of fully  general relativistic  cosmologies and  develop it
into a  precise definition for understanding  the dynamical properties
of  Friedman-Robertson-Walker  (FRW)  spacetimes.  This  framework  is
pedagogically superior  to ostensibly simpler  misleading formulations
of  expanding  space  ---  or  more general  schemes  to  picture  the
expansion  of   the  Universe  ---   such  as  kinematic   models  and
approximations to special relativity  or Newtonian mechanics, since it
is  both clearer  and easier  to understand  as well  as being  a more
accurate approximation. In particular,  it must be emphasised that the
expansion of space  does not, in and of  itself, represent new physics
that is a  cause of observable effects, such  as redshift.  Rather the
expansion  of space is  an intuitive  framework for  understanding the
effects of General Relativity.

In  Sections~\ref{picture} to \ref{analogies}  we detail  this working
picture  of  expanding  space, before  Section~\ref{challengeparticle}
explores this definition  in terms of the motion  of free particles in
an  expanding universe.  The application  of this  approach  to dispel
arguments   against   the   expansion   space   is   demonstrated   in
Section~\ref{common},   before  the   conclusions  are   presented  in
Section~\ref{Conclusion}.

\section{Expanding Space}\label{expandingspace}
In understanding  the concept of expanding space, it is
important to  examine the basic premise of  general relativity, neatly
packaged in John Wheeler's adage
\begin{quote}
matter tells Spacetime how to  curve, and Spacetime tells matter how
to move
\end{quote}
which sets out the dynamical relationship  between the
geometry  of  spacetime  and  the  density and  pressures  of  fluids
contained  therein. 

However, if the prominent cosmologists quoted in the previous section,
will ask ``how can space, which is ultimately empty, expand'', we must
also ask the question of how this ``nothingness'' of the vacuum can be
curved?  By reducing Wheeler's adage to

\begin{quote}
matter tells matter how to move
\end{quote} 
the concept of spacetime, just like the aether, can be banished as
being  non-existent   and  unnecessary  \citep[e.g.][]{Chodconformal}.
Such a  picture is not as  heretical as it seems;  Weinberg (1972), in
his classic text on  general relativity, questions the whole geometric
picture  of  relativity,  and  the  language  it  encompasses,  as  an
unfortunate hangover which is not necessary.

It is enlightening to realise that
 this situation occurs in many
branches  of  physics.  For  example,  in  terms analogue to Wheeler's  adage,
electromagnetism can be reduced to
\begin{quote}
charges tell charges how to move
\end{quote}
but  the employed  framework  contains the  concepts  of electric  and
magnetic   fields   which  are   as   intrinsically  unobservable   as
spacetime.   Furthermore,  these   fields  obey   strict  mathematical
relationships,   through  the   equations  of   Maxwell,   and  many
researchers  can picture the  evolution of  these fields  in dynamical
circumstances, even though it is  just charges telling charges what to
do.

Hence, we  arrive at the view  point that while  general relativity is
just  ``matter telling matter  how to  move'', its  framework contains
deformable  and stretchy  spacetime.  As  with  electromagnetism, this
field  is   not  intrinsically   detectable,  but  does   obey  strict
mathematical relations.  Similarly, as  correct as it  is to  think of
electromagnetism  in terms  of electric  and magnetic  fields,  we can
think  of general  relativity  in  terms of  the  dynamical entity  of
spacetime  as  long  as we  develop  our  intuition  in terms  of  the
underlying  mathematics,  and not  try  to  match  the properties  of
spacetime to the properties of dough  or rubber; just as it would make
no sense to  attempt to construct a physical  or thought experiment to
attempt to prove or disprove  the real existence of magnetic fields, it
is similarly  meaningless to discuss  the expansion of space  in these
terms.

To  illustrate  how short  this  pragmatic  formalism  falls of  being
platitude, one need  look no further than \citet{2006astro.ph.12155A},
in which a  thought experiment of laser ranging in  an FRW Universe is
proposed to `prove' that space  must expand.  This is sensibly refuted
by  \citet{Chodconformal}, but  followed by  a  spurious counter-claim
that such  a refutation  likewise proves space  does not  expand.  The
exercise is futile: what matters  on a technical level are predictions
for observable quantities, which of  course are the same regardless of
how the problem is pictured and what co-ordinate system is chosen. The
expansion of  space is no more  extant than magnetic
fields  are, and  exists only  as  a tool  for understanding  the
unambiguous predictions  of GR, not  a force-like term in  a dynamical
equation. A  recent example  of the dangers  of thinking  of expanding
space  as  a  real  physical   theory  is  contained  in  Table  2  of
\citet{Lieu2007} in  which the expansion  of space is  lumped together
with  the  Big Bang,  Dark  Energy, Dark  Matter  and  Inflation as  a
physical theory demanding verification.

We can certainly agree that this kind  of misuse of the
term ``expansion  of space'' is  fallacious and most  certainly dangerous.
But throwing  the baby of an intuitive framework out
with  the  bathwater  of  misconceptions  leaves  us  only  with  bare
mathematics, which  in the case of general  relativity is particularly
daunting for the uninitiated, and useless as a conceptual device.

\subsection{The Cosmological Picture}\label{picture}
We turn now  to outlining the way in which the  expansion of space can
be  retained  as  a  useful  pedagogical device,  while  avoiding  the
pitfalls of misleading formulations.   It is worth starting from first
principles  and  asking  what  the  general  relativistic  picture  of
cosmology  actually  contains.    The  adoption  of  the  cosmological
principle,  in  that  the   Universe  is  homogeneous  and  isotropic,
restricts  the  form  of  the  underlying geometry  of  the  Universe,
expressed  in  terms  of  the  FRW  metric.   With  this  metric,  the
continuity equation  demonstrates that in  other than finely-tuned or
contrived examples,  the density and pressures  of cosmological fluids
must change over  cosmic time, and it is  this change that represents
the basic property of an expanding (or contracting) universe.

The general relativistic picture  also allows the definition of privileged,
co-moving  observers (said  to reside  in the  Hubble flow)  within the
expanding  universe, those at  rest with  respect to  the cosmological
fluids; in  our Universe, we know  we are not one  of these privileged
observers as our measured CMB  dipole reveals our peculiar motion with
respect to the background photons.   Being at rest with regards to the
cosmic fluid, the proper time  for these privileged observers ticks at
the same rate  as cosmic time and hence the  watches of all privileged
observers are  synchronised. In an  expanding universe, the  change of
the  metric  implies  that  the  physical  distance  between  any  two
privileged observers  increases with time, and  consequently, if eight
of these co-moving observers are used  to define the corners of a cube,
the volume of the cube must increase with time.

Remembering  that  the FRW  metric  describes  a homogeneous  universe
filled  with  a fluid  of  uniform  density,  and assuming  that  test
observers can  measure their velocity  with respect to that  fluid, we
can now describe the formal statement of the phenomenon we refer to as
expanding space:
\begin{quote}
The  distance between  observers at  rest with  respect to  the cosmic
fluid increases with time.
\end{quote}
Since two bodies, both at rest  with respect to the fluid defining the
FRW  metric, find  the distance  between  them has  increased after  a
certain time interval, it seems sensible to suggest that there is more
space between them than there  was previously. It may be misleading to
suggest that the space that was there stretched itself as the universe
expanded.   Perhaps  a better  description,  in  simple  terms, is  to
suggest  that more  space appeared,  or  `welled up'  between the  two
observers, however this is a largely semantic distinction.

We are also in a good  position to understand why the expansion can be
thought  of  locally in  kinematical,  even  Newtonian  terms. We  can
imagine  attaching a  Minkowski  frame  to each  point  in the  Hubble
flow. The local cosmological fluid  is stationary with respect to this
frame.   Whilst only  perfectly accurate  in an  infinitesimally small
region,  the Minkowski  frame  can  be used  as  an approximation  for
regions much smaller  than the Hubble radius. The  Hubble flow is then
viewed as  a purely kinematical phenomenon ---  objects recede because
they   have  been  given  an initial  velocity  proportional   to
distance. This does not argue against expanding space: the equivalence
principle  guarantees  that  any   free-falling  observer  in  any  GR
spacetime  can use  SR locally.\footnote{The  kinematical view  can be
useful, but remains  only a local approximation. The  exception is the
Milne  model:  in   an  empty  universe  we  can   make  a  coordinate
transformation that exchanges the  FRW metric for the Minkowski metric
[see \citet{2000cosm.book.....H},  p.  88], effectively  extending our
local Minkowski frame to all spacetime.  This is only possible because
there  is  no cosmological  fluid  to define  the  rest  frame of  the
universe.   Hence the  Milne  model  cannot be  used  to make  general
comments   on  the   nature   of  the   cosmological  expansion,   cf.
\citet{2006astro.ph..1171C}.    Recently  it   has  been   claimed  by
\citet{Chodconformal}  that conformal  transformations of  general FRW
metrics  can  produce a  common  global  frame  describing the  entire
spacetime, analogous to the common Minkowski frame in the Milne model.
This will  be examined in a  future contribution (Lewis et  al., {\it in
prep}).}

\subsection{Local expansions} \label{localexpansion}
At the global level,  \cite{Peacockweb} suggests that the expansion of
space is uncontroversial since
\begin{quote}
the total  volume of a  closed universe  is a
well-defined quantity that increases with  time, so of course space is
expanding.
\end{quote}
but  questions whether
\begin{quote}
this concept has a  meaningful {\it local} counterpart?\dots Is the space
in my bedroom expanding, and what would this mean?
\end{quote}
Retaining the relativistic  picture of expanding space, it  is easy to
address the question of what happens to Peacock's bedroom, namely
it will  evolve as determined  by the relativistic equations.   But as
ever,  knowledge  of  the   scenario,  and  particularly  the  initial
conditions, is  vital; the walls of  the bedroom are  held together by
electromagnetic forces and hence  are not following geodesics, and the
distribution of  matter has collapsed and  is not uniform,  and so the
underlying  geometry   of  spacetime  in  this  region   needs  to  be
calculated; it  would not be represented  by the FRW  spacetime of the
homogeneous  and isotropic  universe.  Clearly,  if the  universe were
homogeneous  on scales smaller  than Peacock's bedroom,  and the
walls were not  held together by electromagnetic or  other forces, and
the particles  making up the wall  were at rest  with the cosmological
fluid which, importantly, requires that  they not be initially at rest
with respect to  one another, then indeed as  the universe expands the
total  volume of  the  bedroom would  increase.   The many  conditions
listed above are (at least  approximately) true for galaxies not bound
in common groups and hence they  behave in ways that can be understood
and predicted via the framework of expanding space.

This leads to an important point, namely that we should not expect the
global behaviour of a perfectly  homogeneous and isotropic model to be
applicable when  these conditions are not even  approximately met. The
expansion of space fails to  have a `meaningful local counterpart' not
because there is some sleight  of hand involved in considering the two
regimes but because the  physical conditions that manifest the effects
described  as  the expansion  of  space are  not  met  in the  average
suburban bedroom.

\subsection{Dark Energy}\label{darkenergy}
In a matter-dominated universe, the statement in the preceding section
regarding the metric in the  region of a collapsed object being unlike
the  FRW  metric is  straightforward.   However,  if  the universe  is
dominated  (as we  believe ours  currently is)  by an  energy  that by
definition is homogeneous, or  only inhomogeneous on very large scales
then we must be more careful.  In this case the dominant driver of the
specifics   of  the  expansion   rate  will   apply  equally   on  all
scales. However,  so long  as the  equation of state  $w$ of  the dark
energy obeys the condition $w\ge-1$ the energy density will not increase
with  time  and  bound   structures  will  remain  bound  and  stable.
Effectively the region  of spacetime inside a bound  structure will in
fact be matter-dominated, even  though the global mean density is dark
energy-dominated.

\subsection{The Value of Analogies}\label{analogies}
What efficacy then, if any, do the common expanding universe analogies
have?  The balloon-with-dots or bread-with-raisins analogies, like any
analogies,  are  useful  so  long   as  we  are  aware  of  what  they
successfully illustrate  and what constitutes pushing  the analogy too
far.   They show  how a  homogeneous expansion  inevitably  results in
velocity being  proportional to distance, and also  gives an intuition
for how the expansion of the  universe looks the same from every point
in the  universe.  They illustrate  that the universe does  not expand
into previously existing empty  space; it consists of expanding space.
But using these  analogies to visualise a mechanism  like a frictional
or  viscous force  is  taking  the analogy  too  far.  They  correctly
demonstrate the effects of the  expansion of the universe, but not the
mechanism.  That they  fail at some level is  hardly surprising: we're
representing  4-dimensional  pseudo-Riemannian  manifolds  with  party
supplies.  We can't manipulate frames like gravity can.

\subsection{The Challenge of Particle Motion}\label{challengeparticle}
We now turn to the issue of test particle motion, since this is at the
heart of many of the  attacks on expanding space.  The classic thought
experiment    used     is    the    ``tethered     galaxy''    problem
\citep{1995ApJ...446...63H}.  In  this, a test galaxy  in an expanding
universe is held at rest with  respect to the origin at a cosmological
distance.   By  Hubble's  law,  we  would expect  this  galaxy  to  be
receding,  however  we prevent  this,  artificially  holding the  test
galaxy in  place. The question is,  when the galaxy  is released, what
does it  do? Since critics of  the expanding space  concept argue that
the Newtonian analogue  of expanding space is the  presence of some kind
of  viscous force, dragging  the galaxies  apart like  objects carried
along  by a  river: therefore,  in this  thought experiment,  the test
particle should  pick up a  velocity away from  the origin due  to the
expanding river of space.  In  fact, what the particle does once being
released depends  on the acceleration  of the universe.  If  the scale
factor  is  accelerating  the  particle   moves  away  but  if  it  is
decelerating   the   particle    moves   towards   the   origin   [see
\citet{barnes2006}  for  the  full  details].   That  acceleration  is
important in  the question of the  future velocity of a  particle is a
concept that a  student of the most elementary  physics is comfortable
with. Since  expanding space has  apparently mislead our  intuition so
severely   this   appears  to   demonstrate   the   dangers  of   this
interpretation. But  is this a fair  test of expanding  space?  If the
balloon or  baking-bread  analogy is used  to attempt to  picture this
situation  then indeed  the  incorrect answer  is  easily reached.
However, as mentioned, this is  pushing these analogies too far, since
they are  useful in picturing  what an expanding universe  looks like,
but do not speak of what drives that expansion.

This is the central issue  and point of confusion. Galaxies move apart
because they did  in the past, causing the density  of the Universe to
change  and  therefore  altering  the  metric of  spacetime.   We  can
describe this alteration as the  expansion of space, but the key point
is that it is  a result of the change in the  mean energy density, not
the  other way  around.  The  expansion of  space does  not  cause the
distance  between  galaxies  to  increase,  rather  this  increase  in
distance causes space to expand, or more plainly that this increase in
distance is described  by the framework of expanding  space.  There is
therefore no  need to look for  Newtonian analogues to  the expansion of
space, since  it is an effect rather  than a cause.  In  any case, why
should we be seeking Newtonian analogues when we know general relativity
describes the situation well, can be described in simple terms and any
Newtonian   view  will  break   down  at   at  a   non-trivial  limit?
\citet{2004Obs...124..174W} describes  the tethered galaxy  problem in
Newtonian terms and uses Newtonian equations to make predictions about
the asymptotic behaviour  of the test particle.  However,  as shown in
\citet{barnes2006}  these  equations  deviate significantly  from  the
general relativistic results which  begs the question of why Newtonian
analogues should be sought for fundamentally relativistic problems?

Can  the tethered  galaxy  problem  by understood  in  the context  of
expanding space?   We contend that the  answer is yes.  The  key is to
carefully   examine   the   initial   conditions  of   the   particle.
\citet{2004Obs...124..174W} and  \citet{Peacockweb} have in  mind that
the  initial conditions  of the  problem describe  a  particle dropped
innocently into the  universe.  It has no proper  velocity and thus no
prejudice---it is free  to go wherever expanding space  wishes to take
it.  This  is certainly true from a  kinematic, Newtonian perspective:
the particle  is at rest in  our chosen inertial  frame and approaches
the origin due  to the gravitational attraction of  the matter between
the particle and  the origin.  This is locally  valid and even useful,
but it is  not how to understand the scenario  from an expanding space
perspective. The motion of the  particle must be analysed with respect
to its local  rest frame of the test particles,  provided by the Hubble
flow.   In  this  frame,  we  see  the  original  observer  moving  at
$v_{\textrm{rec},0}$  and the particle  shot out  of the  local Hubble
frame at $v_{\textrm{pec},0}$, so  that the scenario resembles a race.
Since their velocities are initially  equal, the winner of the race is
decided by how  these velocities change with time.   In a decelerating
universe, the  recession velocity of the  original observer decreases,
handing  victory to  the  test  particle, which  catches  up with  the
observer.

The   difference   between   the   kinematic   and   expanding   space
interpretation    is    well    illustrated    by    Figure    1    of
\citet{2001astro.ph..4349D}.    Figure    1a   shows   the   kinematic
perspective ---  the observer  and the tethered  particle are  at rest
with respect  to each other,  and gravitational attraction  will bring
them together.   Figure 1b shows the  scenario as seen  from the local
rest frame of the tethered  particle, i.e. a race between the original
observer and the test particle.  The original observer should view the
initial  conditions of the  test particle,  not as  neutral, but  as a
battle between motion  through space and the expansion  of space.  The
expansion  of space  has  been momentarily  nullified  by the  initial
conditions, so  we must  ask how the  expansion of space  changes with
time.

We  contend  that  this  explanation  successfully  incorporates  test
particle motion into the concept of expanding space. In particular, it
shows why it is wrong to  expect, on the basis of the balloon analogy,
that expanding space would carry the particle away. The alternative is
either to  give up on  a physical concept  entirely, so that  the only
rationale for the  cosmological facts is that ``that's  what the maths
tells us'', or to formulate a new framework into which these facts and
more can be  accommodated.  The first option is  unsavoury, the second
unlikely,  unless  one wants  to  discard  GR  entirely and  formulate
cosmology using only Newtonian ideas \citep{1996MNRAS.282..206T}.

\subsection{Using Expanding Space}\label{common}

In this  section, we examine in  detail the employment  of the
concept  of expanding  space in  a number  of  cosmological scenarios.

\subsubsection{Superluminal Recession Velocities}

By failing to place a limit  on the range of validity of Hubble's Law,
the  FRW metric  implies that  there is  no speed  limit  on recession
velocities, seeming  to violate a fundamental  principle of relativity
by implying superluminal motion. This is a frequent cause of concern
and  confusion.   In terms  of  the  proper  distance $D$  defined  as
$D=a\chi$  and  the  cosmic  time  $t$  in the  FRW  metric  then  the
differential  $\frac{dD}{dt}$  most certainly  can  exceed unity,  and
hence by this definition  of velocity, represents superluminal motion.
However,     as    shown     in     \citet{2006astro.ph..3162G}    and
\citet{noSuperLum} we  can, by a  co-ordinate transformation, describe
simple Minkowski  space-time as the  FRW metric of an  empty universe,
known  as the  Milne  model.  In  the  Minkowski special  relativistic
co-ordinates we of course  cannot have superluminal motion. However in
the new Milne model co-ordinates we  do find that $\frac{dD}{dt} > 1 $
beyond a  certain distance  from the origin.  Thus we  have apparently
described  superluminal motion  in  a spacetime  that  we know  cannot
permit  such  a  phenomenon.   Via conformal  transformations,  it  is
possible   (see  e.g.   \citet{Chodconformal})   to  make   a  similar
transformation  between general  FRW metrics  and  conformally related
Minkowski-like metrics.   Again in  the FRW case  $\frac{dD}{dt}$ make
exceed  unity,  while  in  the  conformal co-ordinates  the  speed  is
limited.    While  some   authors   (e.g.  \citet{Chodconformal}   and
\citet{noSuperLum})   have   argued   that  this   demonstrates   that
superluminal   recession   is   impossible,   others,   for   instance
\citet{2006astro.ph..3162G} have argued that superluminal recession is
a  fundamental  consequence of  the  FRW  metric.   As pointed  out  in
\citet{barnes2006},  many of the  debates surrounding  expanding space
turn  out to  be  based  on different  definitions  of poorly  defined
concepts,  in  this  case  the  term  superluminal.   If  we  mean  by
superluminal  that the  motion described  in the  co-ordinates  of the
Minkowski (or conformal Minkowski-like) frame defined by extending the
local inertial frame  of an given observer is  greater than unity then
everyone agrees  that this does  not occur. On  the other hand,  if we
take the  FRW co-ordinates it is clear  that there is no  limit on the
recession velocity:  if we choose  to call this  superluminal motion,
then it indeed occurs.  The debate  seems to boil down to whether this
should or  should not be  given the name `superluminal'  but crucially
the physical predictions  made by either camp will  be identical. What
matters  is  not   what  we  call  the  phenomenon   but  whether  the
understanding an individual  has of a given term  reflects reality and
it  is clear  that not  all the  authors mentioned  above  held common
meanings of the term superluminal.

What does  matter is  that we have  a framework for  understanding the
consequences  of  the  FRW  metric  that  is  unambiguous  and  easy  to
understand. In the seminal work of \citet{2004PASA...21...97D} several
common  mistakes  regarding recession  velocities  are examined.   The
authors  take  a strong  view  that  recession  velocities really  are
superluminal but more importantly show  the types of mistakes that can
be made by  on the one hand  using the FRW formalism and  on the other
hand  making ad  hoc  `corrections' to  prevent apparent  superluminal
motion in  the FRW co-ordinates.  The  key point to take  from this is
that one must be consistent; if  we use the very convenient FRW metric
we  must be  aware that  the  recession speed  is not  limited in  any
way. If this is uncomfortable alternative co-ordinates may be adopted,
and if used consistently will  return the same physical predictions as
the correctly used FRW co-ordinates.

We  will now  outline how  velocities can  be treated  consistently and
clearly  within the  framework of  expanding space.  Consider  a test
particle  moving radially  with  coordinate velocity  $\dot{\chi}(t)$.
The proper  velocity of the object  as measured by an  observer at the
origin is:
\begin{equation}
\dot{r}_\textrm{p} = \dot{R}(t)\chi(t) + R(t)\dot{\chi}(t)
\end{equation}
The first term is the same as for a particle in the Hubble flow at the
same co-moving  coordinate and depends on  the rate of  increase of the
scale factor.  It is  zero for an object at the origin  or in a static
universe. Now, consider the second  term: the time measured on a clock
($\tau$) attached to the particle is given by the FRW line element as
\begin{align}
c^2\textrm{d}\tau^2 &= c^2  \textrm{d}t^2 - R^2(t) \textrm{d}\chi^2 \\
\Rightarrow \quad  \left(\frac{\textrm{d}\tau} {\textrm{d}t} \right)^2
&= 1 - \left( \frac{R(t)\dot{\chi}(t)} {c} \right)^2
\end{align}
Since  $\tau$ is  observable  it must  be  real (zero  for a  photon):
$(\textrm{d}\tau)^2  \ge  0$  implies  that  $|R(t)\dot{\chi}(t)|  \le
c$. Thus, the velocity of the  particle due its motion relative to the
Hubble flow  (or equivalently the  homogeneous fluid defining  the FRW
metric) must be less than the  speed of light; its velocity due to the
increase of the scale factor is not restricted in this way.

We interpret  $\dot{R}(t)\chi(t)$ as the  increase in distance  to the
object due to the expansion of the space between the observer and test
particle (recession velocity), and $R(t)\dot{\chi}(t)$ as the velocity
of object  due to  its motion through  the local rest  frame (peculiar
velocity).   As  previously mentioned,  we  can  consider attaching  a
Minkowski frame to  each point in the Hubble flow.   Then the speed of
light limits the speed of an  object through space. But since there is
no global Minkowski inertial frame  (except for in an empty universe),
the relative  motion of different regions  of the Hubble  flow sees no
speed  limit.   Note that  the  kinematical  view  sees no  difference
between  recession and  peculiar velocities,  and thus  cannot explain
this result.  As an illustration,  for light moving radially away from
the origin: $v_{\textrm{pec}} = c$,  so that $\dot{r}_\textrm{p} = c +
H(t)r_\textrm{p}  > c$.  An  observer who  insists on  extending their
Minkowski frame  into expanding  space will encounter  light travelling
faster than light!

Note that  while the above has ascribed  a velocity to be  `due to the
expansion  of  space',   we  again  stress  that  this   is  a  useful
description,  rather than  a physical  cause or  law.  The  physics in
operation  is  general  relativity  and  the  ultimate  cause  of  the
evolution                of               the               quantities
($r_p,\dot{r_p},\chi,\dot{\chi},R,\dot{R}$)   is  the  characteristics
(summarised by  the equation of  state) and initial conditions  of the
energy in the Universe.

While the  picture of expanding space possesses  distant observers who
are  moving  superluminally, it  is  important  not  to let  classical
commonsense guide your intuition. This would suggest that if you fired
a  photon at  this  distant observer,  it  could never  catch up,  but
integration of  the geodesic equations  can reveal otherwise  (this is
very  clear   in  the   conformal  representation  of   FRW  universes
\citep{Chodconformal}; this  will be examined more deeply  in a future
contribution  (Lewis et  al {\it  in  prep}).  Hence,  again, what  is
important  is   not  the  statement  of   superlumininal  motion,  but
implications for  observations, and these  must be independent  of the
framework in which you choose to work.

\subsubsection{Is everything expanding?}
An extension of the argument against global expansion given in section~\ref{localexpansion}
is that is should be  undetectable, since everything will simply  expand with it.
However, this is not the case: consider a `normal  object', by which  we mean one
consisting  of  many  particles,  held together  by  internal  forces.
Suppose that the centre of  the object travels along a radial geodesic
$\chi_c(t)$ in FRW  spacetime.  Suppose further that the  front of the
object  travels along  a trajectory  $\chi_f(t)$  that keeps  it at  a
constant     proper     distance     ($L$)    from     the     centre,
i.e. \label{normalobject}
\begin{align}
R(t)\chi_f(t) - R(t)\chi_c(t) &= L \qquad \text{(a constant)} \\
\Rightarrow \quad \chi_f(t) &= \chi_c(t) + \frac{L} {R(t)} \label{eq:chifront}
\end{align}
The back  of the object will  move along an analogous  path.  Then the
coordinate trajectory $\chi_f(t)$ is  not a geodesic of FRW spacetime.
The  foremost particle  will  experience a  four-force,  which can  be
calculated  by  substituting  Equation  \eqref{eq:chifront}  into  the
equation of motion of a particle experiencing a four-force $f^a$:
\begin{equation}\label{eq:fourforce}
\frac{\textrm{d}^2 x^a} {\textrm{d} \lambda^2} + \Gamma^a_{bc} 
\frac{\textrm{d}x^b} {\textrm{d}\lambda} \frac{\textrm{d}x^c} 
{\textrm{d}\lambda} = \frac{f^a} {m}
\end{equation}
The  observed force  in the  radial direction  is given  by projecting
$f^1$ onto an  orthonormal basis; the final result  is equation (1) of
\citet{1995ApJ...446...63H} with $U(t) =  -H(t)L$ for all time. In the
case of $L$ small (compared to $c/H$, the Hubble radius), we have that
the radial force $F$ is:
\begin{equation} \label{eq:boundforce}
F = -mL\frac{\ddot{R}} {R}
\end{equation}

This result tells us how  not to understand expanding space. Expanding
space does  not stretch rigid  rulers --- how  could it?  It is  just a
trick with inertial frames.  The internal, interatomic forces in rigid
objects    are   able   to    maintain   the    object's   dimensions;
\citet{1964PhRvL..12..435D}   [see  also  \citet{2006gr.qc.....2098C}]
argue  that EM  forces do  just this.   Objects are  held  together by
forces  that  pull their  extremities  through  a  succession of  rest
frames.

It is worth  considering what would happen to an  object if there were
no electromagnetic forces holding  it together.  Consider an object of
many  particles with  no internal  forces. It  is shot  away  from the
origin ($\chi  = 0$) with speed  $v_0$, the first  particle leaving at
time $t_0$  and the  last at $t_0  + \Delta  t_0$.  The length  of the
object is $l_0 = v_0 \Delta t_0$. The object travels to an observer in
the  Hubble flow at  $\chi$, who  measures its  speed relative  to him
($v_f$) and the time of arrival of the first ($t_f$) and last particle
($t_f + \Delta t_f$) in order to measure its length ($l_f = v_f \Delta
t_f$).  Following \citet{barnes2006}:\label{looseobject}
\begin{equation}
\chi  = \int_{t_0}^{t_f}  \frac{\textrm{d}t} {R\sqrt{1  + C_0  R^2}} =
\int_{t_0+ \Delta t_0}^{t_f + \Delta t_f} \frac{\textrm{d}t} {R\sqrt{1
+ C_f R^2}}
\end{equation}
where $C_0$ and  $C_f$ are calculated from the  initial conditions for
each particle.  If we assume that $\Delta t_0$
and $\Delta t_f$ are small, it follows that we can assume $C_0
=  C_f \equiv  C$ and  then rearrange  the limits  of the  integral to
give\footnote{This part of the derivation is similar to the derivation
of the cosmological redshift directly  from the FRW metric; see, among
many others, \citet{2005gere.book.....H}, p. 368.}
\begin{align}
\int_{t_0}^{t_0+ \Delta t_0} \frac{\textrm{d}t} {R\sqrt{1 + C R^2}} &=
\int_{t_f}^{t_f +  \Delta t_f} \frac{\textrm{d}t} {R\sqrt{1  + C R^2}}
\\ \Rightarrow  \quad \frac{\Delta t_0} {R(t_0)\sqrt{1  + C R^2(t_0)}}
&= \frac{\Delta t_f} {R(t_f)\sqrt{1 + C R^2(t_f)}}
\end{align}
Then,   following the method of \citet{barnes2006}   to  calculate 
  $v_f  = \dot{\chi}(t_f) R(t_f)$ and substituting for $C$ we have that
\begin{equation}
\frac{l_f}  {l_0}   =  \frac{v_f  \Delta  t_f}  {v_0   \Delta  t_0}  =
\frac{R(t_f)} {R(t_0)}
\end{equation}
Hence, the length  of the object has increased  in proportion with the
scale  factor.

This result  answers the question: what  if an object  had no internal
forces, leaving it  at the mercy of expanding space?  This is a rather
strange object --- it would very quickly be disrupted by the forces of
everyday life.   Nevertheless, it is a useful  thought experiment. The
above result  shows that the  object, being subject only  to expanding
space, has been stretched in  proportion with the scale factor.  These
are essentially cosmological tidal forces.

We therefore have clear,  unambiguous conditions that determine whether
an object  will be stretched by  the expansion of  space. Objects will
not expand with the universe when there are sufficient internal forces
to maintain the  dimensions of the object.

\subsubsection{Why aren't galaxies or clusters pulled apart by the
  expansion of space?}\label{propstretch}

Having dealt with  objects that are held together  by internal forces,
we now  turn to objects  held together by gravitational  `force'.  One
response to the question of  galaxies and expansion is that their self
gravity is  sufficient to  `overcome' the global  expansion.  However,
this suggests  that on the  one hand we  have the global  expansion of
space acting as the cause, driving matter apart, and on the other hand
we  have gravity  fighting  this expansion.   This hybrid  explanation
treats gravity  globally in general relativistic terms  and locally as
Newtonian,  or  at best  a  four force  tacked  onto  the FRW  metric.
Unsurprisingly then, the resulting picture the student comes away with
is  is  somewhat murky  and  incoherent,  with  the expansion  of  the
Universe having mystical properties.   A clearer explanation is simply
that on  the scales  of galaxies the  cosmological principle  does not
hold, even approximately, and the  FRW metric is not valid. The metric
of spacetime  in the region  of a galaxy  (if it could  be calculated)
would look  much more Schwarzchildian  than FRW like, though  the true
metric would be  some kind of chimera of both.   There is no expansion
for the galaxy to overcome, since the metric of the local universe has
already  been altered  by  the presence  of  the mass  of the  galaxy.
Treating gravity as a four-force and something that warps spacetime in
the one conceptual  model is bound to cause  student more trouble than
the explanation  is worth.  The expansion  of space is  global but not
universal,  since  we  know the  FRW  metric  is  only a  large  scale
approximation.

\subsubsection{The Expansion of Space and Redshift}

The explanation  of redshift is a  crucial link that needs  to be made
between  cosmological observations  and theory.   A derivation  of the
balloon analogy is  often employed in the teaching  of this concept; a
wave is sketched on a balloon and  as it is blown up the wavelength is
seen to increase  as the sketch is stretched  along with the expansion
of the  underlying space.  This  is largely uncontroversial,  but care
must be taken in ensuring that  the analogy does not mislead. Since we
have shown how  bodies held together by electromagnetic  forces do not
expand with  the expansion of space, why  should electromagnetic waves
be affected? The key is to make it clear that cosmological redshift is
not, as is  often implied, a gradual process  caused by the stretching
of  the space  a  photon is  travelling  through. Rather  cosmological
redshift is caused  by the photon being observed  in a different frame
to that which it is emitted. In this way it is not as dissimilar to a
Doppler  shift as  is often  implied.  The  difference  between frames
relates  to  a changing  background  metric  rather  than a  differing
velocity.   Page   367  of  \citet{2005gere.book.....H}   as  well  as
innumerable other texts  shows how redshift can be  derived very simply
by considering the  change in the orthonormal basis  of observers with
different scale factors in  their background metrics.  This process is
discreet, occurring  at the  point of reception  of the  photon, rather
than being continuous, which would  require an integral. If we consider
a series of co-moving observers, then they effectively see the wave as
being stretched with the scale factor.

\section{Conclusion}\label{Conclusion}

Despite (and perhaps in part because of) its  ubiquity, the
 concept of expanding space  has  often been  articulated poorly
 and formulated  in contradictory  ways. That addressing this issue
 is important must be placed beyond doubt, as the  phrase `expansion
 of space' is in such  a wide  use---from technical  papers, through
 to  textbooks and  material intended  for school  students or the
 general public---that it is no exaggeration to label it the most 
 prominent feature of Big Bang cosmologies. In this paper, we have 
 shown how a consistent description of cosmological dynamics emerges
 from the idea that the expansion of  space is neither more  nor
 less  than the increase over time of the distance between  observers
 at rest  with respect  to  the cosmic fluid.  
 
This description of the cosmic expansion  should be  considered a
 teaching and  conceptual aid, rather  than a physical theory  with an
 attendant clutch of physical predictions. We  have demonstrated  the power of
 this pragmatic conceptualisation in guiding  understanding of
 the  universe, particularly in avoiding the traps into which we can  be  lead 
 without rigorous  recourse to general  relativity. 
 
The utility of approximation in handling the less tractable properties of
 cosmologies is undiminished, but the understanding of physical systems therein
 will be hampered wherever full covariance is absent. All observational
 properties---whether derived  in the dynamically  evolving FRW metric
 or  the Minkowski-like conformal  representation---must be  the same,
 independent  of the  choice of  co-ordinates.  As  general relativity
 approaches  its one-hundredth birthday,  this is  a lesson  that all
 cosmologists should learn

\section*{Acknowledgements}

LAB and JBJ are supported  by an Australian Postgraduate Award. LAB is
also supported  by a  University of Sydney  School of  Physics Denison
Merit Award.  JBJ is  also generously supported  by the  University of
Edinburgh.  MJF is  supported by  a  University of  Sydney Faculty  of
Science UPA scholarship.   GFL gratefully acknowledges support through
the  University of  Cambridge Institute of  Astronomy  visitor program. Brendon Brewer is thanked for helping suggest the title.
This work has been supported  by the Australian Research Council under
grant DP  0665574.

\end{document}